\documentclass[numprefix,final,5p,times,twocolumn,sort&compress]{elsarticle}

\usepackage{amssymb}
\usepackage{lipsum}
\usepackage{amsmath}
\usepackage[dvipsnames]{xcolor}
\usepackage{graphicx}
\usepackage{hyperref}
\usepackage{mathtools}
\usepackage{comment}
\usepackage{cancel}
\usepackage{physics}
\usepackage{mathtools}
\usepackage{adjustbox}
\usepackage{multirow,array,booktabs}
\usepackage[shortlabels]{enumitem}
\usepackage{enumitem} 
\usepackage[caption=false,justification=justified]{subfig} 
\newcommand{\jy}[1]{{\color{Green}{#1}}}

\journal{Physics Letters B}

\begin{document}

\begin{frontmatter}



\title{BCS superconductivity in the presence of wave dark matter}


\author[KAIST]{YeolLin ChoeJo}
\ead{particlephysics@kaist.ac.kr}
\author[KAIST]{Yechan Kim}
\ead{cj7801@kaist.ac.kr}
\author[KAIST]{Hye-Sung Lee}
\ead{hyesung.lee@kaist.ac.kr}
\author[KAIST,KIAS]{Jiheon Lee}
\ead{anffl0101@kias.re.kr}
\author[KAIST,KSA]{Jaeok Yi}
\ead{jaeok.yi@ksa.kaist.ac.kr}
\affiliation[KAIST]{organization={Department of Physics, Korea Advanced Institute of Science and Technology (KAIST)},
            addressline={291 Daehak-ro}, 
            city={Yuseong-gu},
            postcode={34141}, 
            state={Daejeon},
            country={Korea}}
\affiliation[KIAS]{organization={Quantum Universe Center, Korea Institute for Advanced Study (KIAS)},
            addressline={85 Hoegi-ro}, 
            city={Dongdaemun-gu},
            postcode={02455}, 
            state={Seoul},
            country={Korea}}
\affiliation[KSA]{organization={Department of Physics \& Earth Science, Korea Science Academy of KAIST},
            addressline={105-47 Baekyanggwanmun-ro}, 
            city={Busanjin-gu},
            postcode={47162}, 
            state={Busan},
            country={Korea}}

\begin{abstract}
As particle physics and condensed matter physics continue to expand their boundaries, emerging intersections between the two fields suggest well-motivated frontiers for physics models. In particular, dark matter, which is cosmologically predicted to form the background to our ordinary matter, can affect the physics of materials in a fundamental sense. In this work, we investigate the scenario in which relic ultralight wave dark matter alters the superconductor Hamiltonian via a sinusoidal source term. The Yukawa interaction with the scalar dark matter changes the effective electron mass, inducing periodic perturbations of macroscopic observables such as the superconducting order parameter, whose size could be important in future experiments. We provide here the actual calculations for this scenario on the superconducting resonators, calculating the modulation size depending on the dark matter parameters.
\end{abstract}




\end{frontmatter}




\section{Introduction}
\label{introduction}

Ever since the establishment of condensed matter physics, our understanding of the theory has developed at a remarkable pace. Grounded in the quantum electromagnetic behaviors of matter constituents, it has brought irreversible changes to our daily lives through various applications. While the physics of solid materials still holds profound mysteries, modern solid-state experiments have reached outstanding levels of sensitivity driven by the need to uncover these dynamics, and are continuously pushing the physical limits of macroscopic measurements.

Meanwhile, on the side of particle physics and cosmology, the largely successful $\Lambda$CDM framework predicts that there exists a dark matter (DM) forming a universal background to our visible matter \cite{Planck:2018vyg}. This halo of DM must alter our conventional viewpoints of physics models, although the assumed feeble interaction strength between the DM and the Standard Model particles is thought to render its impacts hidden until today.

As these two branches of physics continue to extend their boundaries, their meeting generates a promising frontier for the area of novel physics. Although the links between the two have already been substantially explored \cite{Nakonieczny:2015ela, Rogatko:2016cdb, Iwazaki:2020zer, Yao:2025qmi}, most attention until now has been paid to the role of condensed matter devices primarily as DM particle detectors. However, the DM interaction can fundamentally alter the mechanics of the solid state itself, and the profound nature of this under the Beyond Standard Model (BSM) framework stands as a highly intriguing avenue for investigations.

The effects of this intersection can appear differently in various scenarios. One particularly interesting branch is the fuzzy DM, or so-called the `wave DM' \cite{Hui:2021tkt} regime, and it is the case where the DM supplies a coherent background drive to ordinary matter. These ultralight bosonic DM candidates exhibit a large occupation number and thus behave as an oscillating semiclassical field in the current universe. This perspective suggests that a condensed matter system should actually be interpreted as being immersed in a sinusoidally modulating environment.

\begin{figure}
	\centering 
	\includegraphics[width=0.49\textwidth]{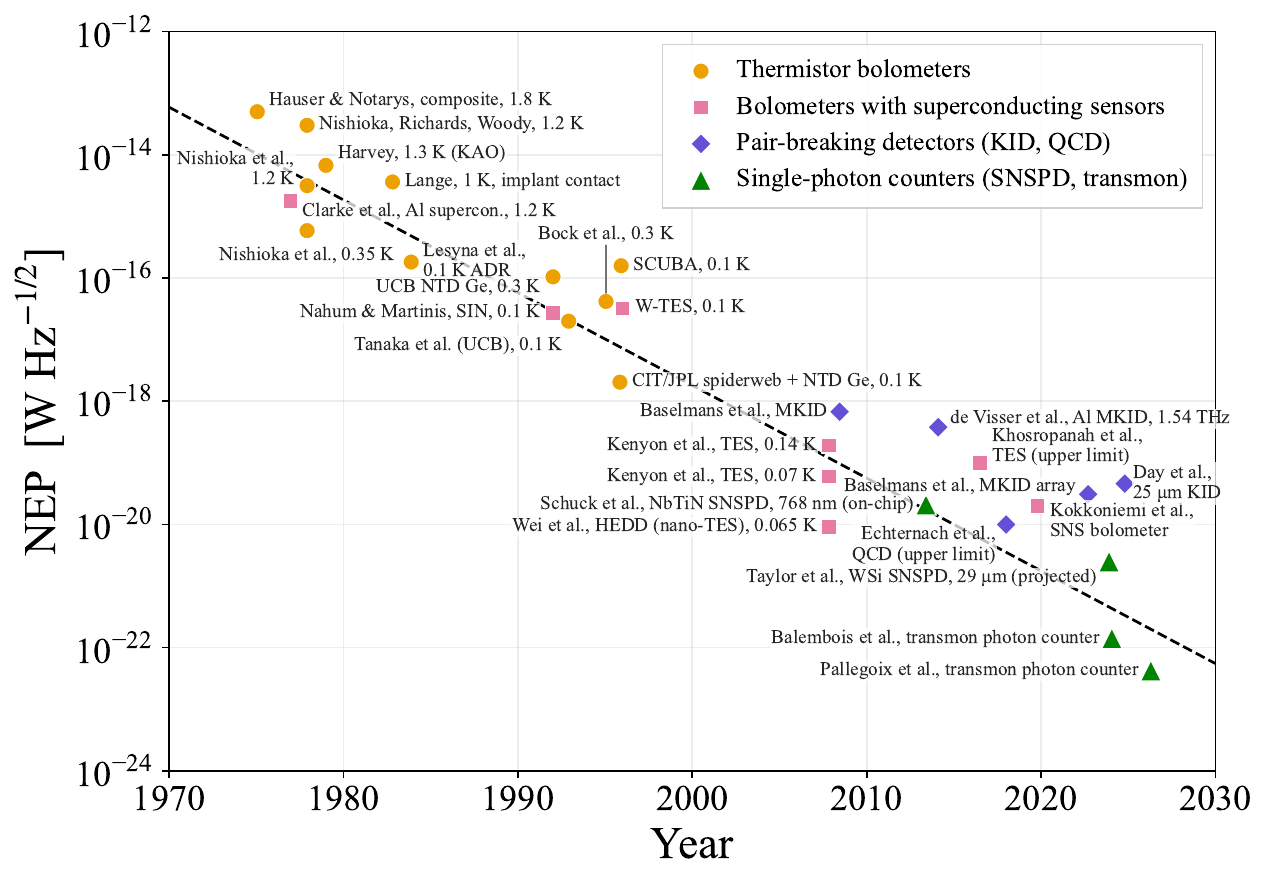}	
	\caption{The schematic figure of the development of noise equivalent power (NEP) of some representative condensed matter experiments \cite{EPIC:2009cnt,Schuck2013,deVisser2014,Khosropanah2016,Echternach2018,Kokkoniemi2019,Baselmans2022,Taylor2023,Balembois2024,Day2024,Pallegoix2026}. Points until 2009 are adapted from Figure 4.3 of \cite{EPIC:2009cnt}. As the model gets more accurate and the experimental techniques get improved, solid-state materials always have the probability of encountering the DM sector. The effect would become significant if the systematic and statistical noises become comparable to the DM interaction strength.
    }
	\label{fig:schematic}
\end{figure}

Figure~\ref{fig:schematic} presents a schematic plot showing the developments of condensed matter experiments. As the sensitivity increases continuously, they would eventually reach the interacting DM background, or equivalently the perturbations generated by the DM halo, requiring new models that properly incorporate the BSM interactions in condensed matter phenomenology.

In the following work, we treat the wave DM as a small but irreducible background source term in condensed matter systems, focusing on the case of superconductors, which share a profound theoretical connection with particle physics. Based on the Bardeen–Cooper–Schrieffer (BCS) theory \cite{Bardeen:1957kj, Bardeen:1957mv}, one can investigate how the universal background of coherent scalar DM modifies the adiabatic BCS solution itself, and finally affects the macroscopic parameters. Beyond the theoretical derivations, we also provide an explicit framework connecting it to actual superconductor experiments.

The BCS system in the presence of a periodic source term was previously studied in Ref.~\cite{roy2013periodic}. However, we find that the calculation there contains an error, which leads to an overestimate of the effect. In this work, we present a correct derivation of this effect for a BCS system with a time-dependent electron interaction term.


This article is organized as follows. In Section~\ref{sec:waveDMInt}, we establish the model for the interaction between wave DM and electrons, and review the current constraints on this Yukawa coupling strength. Section~\ref{sec:BCSwaveDM} details the analysis using a conventional superconductor as a reference system, and constructs the extended BCS theory in the presence of this interaction, where the modification is discussed in the language of modulation of model parameters. Section~\ref{sec:experiment} provides practical estimates about the possible impacts of this modulation in real physical systems, comparing the possible DM interaction strength with the current noise budget of the model system: kinetic inductance detectors. Lastly, we summarize our results and discuss future outlooks in Section~\ref{sec:summary}.


\section{Interaction between wave dark matter and electrons}
\label{sec:waveDMInt}

The interaction of wave dark matter with condensed matter systems is highly model-dependent.
In this work, we consider an effective Yukawa coupling between an ultralight scalar field $\Phi$ and the electron Dirac field $\psi_e$, as can arise in light scalar dark matter scenarios, including dilaton-like models, given by
$\mathcal L \supset - \kappa \Phi \bar\psi_e \psi_e$, which we adopt here as a minimal and representative parametrization.
Although the resulting coupling is known to be stringently constrained, interactions of this form have been widely studied in the literature~\citep{Damour:2010rp, Arvanitaki:2016fyj, Berge:2017ovy, Hees:2018fpg, Kennedy:2020bac, Zhang:2022ewz, Kobayashi:2022vsf, Bottaro:2023gep, Filzinger:2023qqh, Flambaum:2024zyt}.

At the level of the Standard Model, this effective interaction can arise from gauge-invariant higher-dimensional operators, e.g.,
\begin{equation}
\mathcal{L} \supset -\frac{c_e}{\Lambda}\,\Phi\,\bar{L}\,\Phi_H\, e_R + \text{h.c.},
\end{equation}
where \( c_e \) is a dimensionless coefficient, \( \Lambda \) denotes the ultraviolet cutoff scale, \( \Phi_H \) is the Higgs doublet, and \( L = (\nu_L,\, e_L)^T \) is the lepton doublet. After electroweak symmetry breaking, with
\( \langle \Phi_H \rangle = (0,\, v/\sqrt{2})^T \), this operator generates the effective coupling $\kappa = \frac{c_e v}{\sqrt{2} \Lambda}$. 

In the ultralight mass range from $10^{-22}\,\mathrm{eV}$ to $30\,\mathrm{eV}$, the de Broglie wavelength exceeds the interparticle spacing, and the field is well described as a classical wave, which is referred to as `wave dark matter'~\cite{Hui:2021tkt}.
Treating $\Phi$ as a classical field in an expanding universe, its equation of motion is
\begin{align}
\ddot{\Phi} -\nabla^2\Phi + 3H\dot{\Phi} + M_\Phi^2 \Phi +\kappa\langle\bar\psi_e\psi_e\rangle 
= 0,
\label{eq:EOMPhi}
\end{align}
with Hubble parameter $H$ and the DM mass $M_\Phi$. The last term represents the backreaction of electrons on wave DM and is negligible in the parameter space of interest.

Then, in the present epoch, the Hubble friction term is negligible, and the scalar field oscillates coherently as
\begin{align}
\Phi(t) \simeq \frac{\sqrt{2\rho_\Phi}}{M_\Phi}\sin(M_\Phi t),
\label{eq:LocalPhi}
\end{align}
where $\rho_\Phi\simeq 0.4~\mathrm{GeV/cm^3}$~\cite{Bovy:2012tw,McKee:2015hwa,Sivertsson:2017rkp} denotes the local DM energy density.
The interaction with wave DM can induce the time-varying properties of the various particles~\cite{Arvanitaki:2009fg, Damour:2010rp, VanTilburg:2015oza, Berlin:2016woy, Zhao:2017wmo, Krnjaic:2017zlz, Brdar:2017kbt, Dev:2022bae, Guo:2022vxr, ChoeJo:2023ffp, ChoeJo:2023cnx, ChoeJo:2024wqr, Kim:2025xum, 
Arvanitaki:2016fyj, Berge:2017ovy, Hees:2018fpg, Kennedy:2020bac, Zhang:2022ewz, Kobayashi:2022vsf, Bottaro:2023gep, Filzinger:2023qqh, Flambaum:2024zyt}. 
This induces oscillations in electron sector quantities, e.g., an effective $m_e$ shift.

\begin{figure}[tb]
    \centering
    \includegraphics[width=1.0\linewidth]{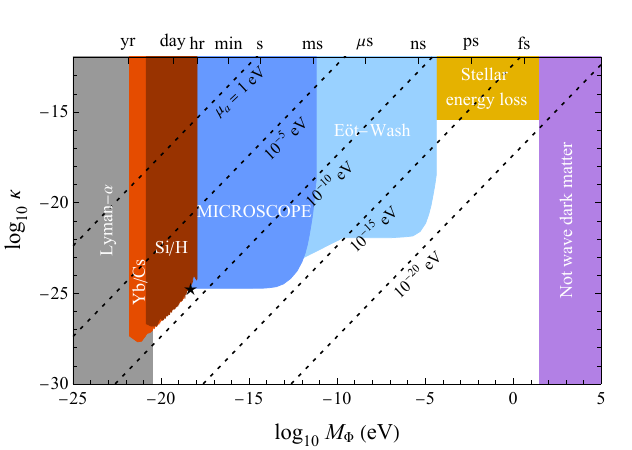}
    \caption{Constraints on the coupling $\kappa$ between wave DM and electrons over $M_\Phi$. 
    Shaded regions indicate atomic clock/cavity, equivalence principle, and stellar energy loss bounds, along with the Lyman-$\alpha$ limit.
    The black dashed lines show the magnitude of induced mass change $\kappa\Phi$, which corresponds to the size of change in effective mass of the electrons.
    The star denotes the benchmark point we used.
}
    \label{constraint}
\end{figure}
Experimental and observational data place stringent bounds on the Yukawa coupling $\kappa$ to electrons. (i) Atomic clocks/cavities for $M_\Phi\sim 10^{-22}\,\mbox{--}\,10^{-17}\,\mathrm{eV}$:
\(\kappa \lesssim 10^{-28}\,\big(M_\Phi/10^{-22}\,\mathrm{eV}\big)\) from frequency modulations~\cite{Kennedy:2020bac, Zhang:2022ewz, Kobayashi:2022vsf, Filzinger:2023qqh}.
(ii) Equivalence principle and fifth force tests:
MICROSCOPE yields $\kappa \lesssim 10^{-25}$ for $M_\Phi \lesssim 10^{-12}\,\mathrm{eV}$~\cite{Berge:2017ovy, Hees:2018fpg, Touboul:2017grn}\footnote{Constraints on $L$ and $B-L$ couplings from MICROSCOPE have also been studied~\cite{Fayet:2017pdp, Fayet:2018cjy, Fayet:2025brx}.}; E\"ot-Wash torsion-balance data imply $\kappa \lesssim 10^{-22}$ for $M_\Phi \lesssim 10^{-5}\,\mathrm{eV}$~\cite{Wagner:2012ui}.
(iii) Stellar energy loss: for $M_\Phi \lesssim 1\,\mathrm{keV}$, emission constraints give $\kappa \lesssim 10^{-15}$~\cite{Bottaro:2023gep, Flambaum:2024zyt}.
These bounds are summarized in Fig.~\ref{constraint}.

\section{BCS theory in the presence of the wave DM}
\label{sec:BCSwaveDM}

Now we consider the effects of this Yukawa interaction term in the superconducting system. One starts with the simplest model, the BCS theory for phonon-mediated superconductors. The standard BCS Hamiltonian is written as
\begin{equation}
    \sum_{\bf{k}, \sigma} \xi_{\mathbf{k}} \hat{\chi}^\dagger_{\bf{k}\sigma} \hat{\chi}_{\bf{k}\sigma} - g \sum_{\bf{k}',\bf{k}} 
\hat{\chi}^\dagger_{\bf{k}'\uparrow} \hat{\chi}^\dagger_{\bf{-k}'\downarrow} \hat{\chi}_{\bf{-k}\downarrow} \hat{\chi}_{\bf{k}\uparrow}.
\label{eq:BCSH}
\end{equation}
in which $\xi_{\mathbf{k}}\equiv\epsilon_{\mathbf{k}}-\mu$ for given electron dispersion $\epsilon_{\mathbf{k}}$ and chemical potential $\mu$, with coupling strength $g > 0$ representing the effective attractive interaction conventionally mediated by phonons. This interaction is non-vanishing only within the energy window $\abs{\xi_{\mathbf{k}}} < \omega_D$, where $\omega_D$ is the Debye frequency.

One can write the eigenstate of this Hamiltonian as
\begin{equation}
    \ket{\psi(t)} = \prod_{\mathbf k} \big(u_{\mathbf k }(t) + v_{\mathbf k}(t) \hat\chi_{\mathbf k \uparrow}^\dagger \hat \chi_{-\mathbf {k} \downarrow}^\dagger \big) \ket{0},
\end{equation}
where the amplitudes $u_{\mathbf k}$ and $v_{\mathbf k}$ satisfy the following matrix equation under the above Hamiltonian, analogous to the Bogoliubov-de Gennes equation: 
\begin{equation}
i \partial_t
\begin{pmatrix}
u_{\bf{k}}(t) \\ v_{\bf{k}}(t)
\end{pmatrix} = \begin{pmatrix}
\xi_{\bf{k}} & \Delta^*
\\
\Delta & -\xi_{\bf {k}}
\end{pmatrix}
\begin{pmatrix}
u_{\bf{k}}(t) \\ v_{\bf{k}}(t)
\end{pmatrix}
\label{eigen},
\end{equation}
with the energy gap $\Delta$ serving as the order parameter of the modified BCS system, defined as\footnote{This relation holds only for $s$-wave pairing, which will be the focus of our discussion. For other types of pairing, such as $d$-wave, additional $\mathbf k$-dependence may arise.}
\begin{align}
\Delta = g \sum_{\bf{k'}}  \langle \hat{\chi}_{\bf{-k'}\downarrow} \hat{\chi}_{\bf{k'}\uparrow}\rangle ,
\label{gapdef}
\end{align}
where $\langle \, \cdots \rangle$ means the average over thermal distribution. The energy gap suggested in Eq. \eqref{gapdef} should satisfy the following ground state self-consistency condition,
\begin{equation}
    \Delta = g \sum_{\mathbf k} u^{\text{ad}~*}_{\bf k}(t) v^{\text{ad}}_{\bf k }(t) = g \sum_{\mathbf k} \frac{\Delta}{2E_{\bf k}}.
\label{eq:selfcons}
\end{equation}

In the adiabatic limit, the solution of this equation is given by
\begin{equation}
\begin{split}
u_{\bf{k}}^{\text{ad}}(t) & = \frac{1}{\sqrt{2}} \left( 1 + \frac{\xi_{\bf{k}}}{E_{\bf{k}}} \right)^{\frac{1}{2}} e^{-i E_{\mathbf k} t}, 
\\
v_{\bf{k}}^{\text{ad}}(t) & = \frac{1}{\sqrt{2}} \left( 1 - \frac{\xi_{\bf{k}}}{E_{\bf{k}}} \right)^{\frac{1}{2}} e^{-i E_{\mathbf k} t}, 
\end{split}
\label{eq:ead}
\end{equation}where $E_{\textbf k} =  \sqrt{\xi_{\textbf k}^2 + \Delta^2}$. 

Now we want to analyze the time-dependent perturbation with respect to these adiabatic solutions. Then the parameters $\xi_{\mathbf{k}}(t)$, $\Delta(t)$ and $E_{\mathbf{k}}(t)$ all become time-dependent. In the presence of wave DM $\Phi(t)$ and its Yukawa interaction with electrons, the modulation originates from the effective shift of the electron mass.\footnote{The interaction between wave DM $\Phi$ and electrons can, in principle, mediate an additional attractive interaction, but this effect is absorbed into the effective parameter $g$, determined experimentally.
}
\begin{equation}
m_e \rightarrow m_e + \kappa\Phi = (1+\eta_m)m_e,
\end{equation}
where $\eta_m \equiv \kappa\Phi/m_e$ is the small oscillating parameter representing the contribution of the DM scalar to the mass.

In the lattice, $m_e \rightarrow m_e'= (1+\eta_m)m_e$ changes the lattice spacing as $a \rightarrow a'= (1+\eta_m)^{-1}a$ and results in a dilation of the electronic momenta as $k \rightarrow k'= (1+\eta_m)k$. Using these relations, the modified electron dispersion reads as
\begin{equation}
\begin{aligned}
\epsilon_{\mathbf{k}}' = \frac{k'^2}{2m_e'} = (1+\eta_m)\frac{k^2}{2m_e} = (1+\eta_m) \epsilon_{\mathbf{k}} = \epsilon_{\mathbf{k}} + \delta\epsilon_{\mathbf{k}},
\end{aligned}
\end{equation}
Assuming $T=0$, denoting the unperturbed Fermi energy as $E_F = k_F^2/2m_e$, the original chemical potential was $\mu = E_F$. The new chemical potential $\mu'=\epsilon'_{k'_F}$ is then
\begin{equation}
\begin{aligned}
\mu' = \frac{k_F'^2}{2m'_e} = (1+\eta_m)\frac{k_F^2}{2m_e} = (1+\eta_m)\mu = \mu + \delta\mu,
\end{aligned}
\end{equation}
resulting in
\begin{equation}
\begin{aligned}
\delta\xi_{\mathbf{k}} = \delta\epsilon_{\mathbf{k}} - \delta\mu = \eta_m\left(\frac{k^2}{2m_e}-\mu\right) = \eta_m\xi_{\mathbf{k}},
\end{aligned}
\end{equation}
which is proportional to $\xi_{\mathbf{k}}$.

The adiabatic assumption nearly holds here, as the modulation is very small and slow, 
and we shall expand this small perturbation from the original adiabatic relations. In the static limit $\kappa \rightarrow 0$, $\eta_m \rightarrow 0$, the solution converges exactly to the standard BCS eigenstates.

We analyze the variation of the gap from the initial non-perturbed adiabatic value given by the self-consistency condition, expanding the energy gap as
\begin{equation}
\Delta(t) = \Delta_0 + \delta\Delta(t) \equiv \Delta_0\big(1+\eta_\Delta(t)\big),
\end{equation}
where the subscript $_0$ denotes the non-perturbed value.

To derive the quantitative relations between $\eta_m$ and $\eta_\Delta$, one first substitutes this perturbed gap into the self-consistency formula Eq. (\ref{eq:selfcons}). Replacing the sum with an integral involving the density of states up to the Debye frequency $\omega_D$,
\begin{equation}
\sum_{\mathbf k }  \to \int_{-\omega_D}^{\omega_D} d\xi_{\mathbf k } \ n(\xi_{\mathbf k}+\mu),
\end{equation}
the equation now becomes
\begin{equation}
1 = g\int_{-\omega_D}^{\omega_D} \frac{d\xi_{\mathbf{k}}\,n(\xi_\mathbf{k}+\mu)}{2\sqrt{\xi_\mathbf{k}^2+\Delta^2}}.
\end{equation}
Since the chemical potential closely approximates the Fermi energy $E_F$ and $\omega_D \ll E_F$,  applying a Taylor expansion to the density of states $n(\xi_{\mathbf k}+\mu)  \approx n(E_F) + n'(E_F)\xi_{\mathbf k}+\mathcal{O}(\xi^2)$, only the $n(E_F)$ term survives at the leading order by parity, and one gets
\begin{equation}
1 = g \cdot n(E_F) \int_0^{\omega_D} \frac{d\xi_{\mathbf{k}}}{\sqrt{\xi_{\mathbf{k}}^2+\Delta^2}},
\end{equation}
at $T=0$.

Denoting $\xi_{\mathbf{k}}(t) = \big(1+\eta_m(t)\big)\xi_{0\mathbf{k}}$
and assuming that all parameters other than $\Delta$ remain fixed,
we expand the gap consistency condition to leading order in the
deviations and obtain
\begin{equation}
\begin{aligned}
\eta_m\int_0^{\omega_D} d\xi_{\mathbf{k}0} ~ \xi_{\mathbf{k}0}^2 (\xi_{\mathbf{k}0}^2+\Delta_0^2)^{-\frac{3}{2}} = - \eta_\Delta\int_0^{\omega_D} d\xi_{\mathbf{k}0} ~ \Delta_0^2 (\xi_{\mathbf{k}0}^2+\Delta_0^2)^{-\frac{3}{2}} .
\end{aligned}
\end{equation}

By evaluating the integrals, one obtains the response of the zero temperature BCS gap to the scaling of electronic mass as
\begin{equation}
\eta_\Delta(t) = -\left(\frac{1}{x}\tanh^{-1}x - 1\right)\eta_m(t) \equiv -\mathcal{D}(x)\,\eta_m(t).
\label{gapmod}
\end{equation}
where $x = \omega_D/\sqrt{\Delta_0^2 + \omega_D^2}$, which is typically very close to 1 as conventional metallic materials have $\Delta_0 \ll \omega_D$. The dimensionless proportionality coefficient $\mathcal{D}(x)$ is given for each material and is typically between 1 and 10, which is $5.0$ for the case of aluminum.

Therefore, one finds that the amplitude of modulation of the superconducting gap is directly proportional to that of the mass oscillation generated via the Yukawa interaction with scalar DM. This indicates that the superconducting gap can serve as a direct and universal window to the Yukawa coupling of the Standard Model particles to the dark sector.

Given that the oscillation amplitude of the DM is determined by the mass and the energy density of wave DM, the amplitude of $\eta_\Delta(t)$ is then calculated as
\begin{align}
\eta_{\Delta,\text{max}} \simeq 10^{-14}\cdot \left(\frac{\mathcal{D}(x)}{5.15}\right) \left( \frac{\kappa}{2 \times 10^{-25}} \right) \left( \frac{5\times 10^{-19}\,\mathrm{eV}}{M_\Phi} \right),
\label{eq:eta_max}
\end{align}
translating the upper bound of wave DM interaction strength into the gap modulation sensitivity. In other words, Yukawa interaction with wave dark matter can give a fractional gap modulation of approximately $10^{-14}$ relative to the static value under current DM coupling constraints.

These predicted order parameter oscillations in condensed matter systems subject to the Yukawa interactions can be analyzed complementarily alongside other methods such as mechanical oscillators \cite{Campbell:2021mnu}. To assess the practical impacts of this gap modulation, one can monitor other physical parameters intimately coupled to the energy gap. When the time-dependent gap $\Delta(t)$ translates into macroscopic quantities, the BCS system is expected to exhibit distinct temporal variations~\cite{roy2013periodic} and as the precision of the gap increases, it should be sensitive to the oscillating background fluctuations generated by the relic wave DM.

\section{DM induced modulations and their experimental impacts}
\label{sec:experiment}

To explore the practical implications of this oscillation by wave DM, we present a model analysis comparing this modulation size with the precision of the actual experiments. Here we consider the kinetic inductance detectors (KIDs), which consist of metallic superconducting thin films~\cite{day2003broadband,mazin2005microwave,gao2007noise,gao2008experimental,gao2008semiempirical,gao2008physics,zmuidzinas2012superconducting,mazin2010thin,ji2014design,beldi2019high,cothard2024parallel,doyle2008lumped,defrance2024characterization,martin2024hierarchical,golwala2024new,defrance2025low,hempel2026low,Golwala:2012dk,Smith:2024ljo}. At low temperatures, the kinetic inductance of these superconductors allows for precise measurements of the system's resonance frequency.

In this scenario, the oscillating potential modulates the superconducting gap, driving a periodic change in the kinetic inductance that directly shifts the resonance frequency. Thus, one can expect that the KID, with sufficient precision in tracking the fractional change in the resonance frequency, could be explicitly affected by the oscillatory DM background.

\begin{figure}[tb]
\centering
\includegraphics[width=0.49\textwidth]{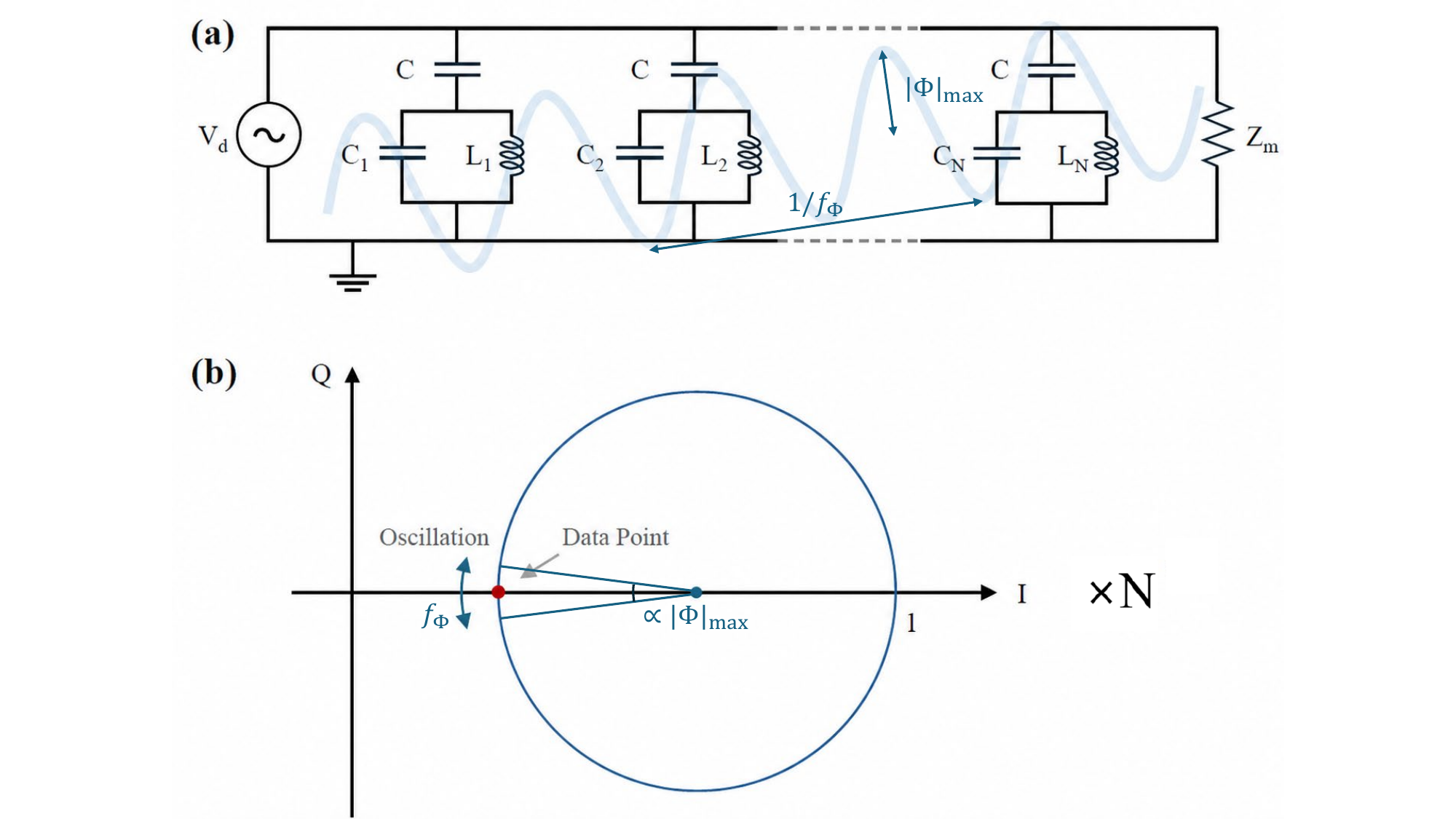}
\caption{
(a) An example diagram of a KID array, submerged in the oscillating background potential from wave DM with the amplitude $|\Phi|_{\rm max}$ and the frequency $f_\Phi$. 
$L_i$ and $C_i$ represent inductance and capacitance, respectively, and $C$ is the common capacitance across the resonators.
$V_d$ is the driving voltage, and the signal is measured at $Z_m$ as a complex amplitude.
(b) In the presence of the DM modulations, the resulting data point oscillates with the same frequency along the circular trajectory in the $IQ$-plane for each $N$ resonators.
Here, $I$ means \textit{in-phase} and $Q$ means \textit{quadrature} component of the complex amplitude.
}
\label{SetupPlot}
\end{figure}
Figure~\ref{SetupPlot} illustrates our model setup of KID arrays submerged in the scalar DM backgrounds. The typical KID readout from the resonator is a complex number, or equivalently, a point on the $IQ$-plane. Scanning of the resonance frequency gives the circular trajectory on the $IQ$-plane, and the frequency is mapped on the circular perimeter. Without background, once the driving frequency is fixed, the data point representing the complex amplitude of the transmitted wave should be anchored on the plane.

In the presence of an oscillating perturbation on the resonance frequency, the change in this frequency will appear as a movement of the point~\cite{gao2008physics,zmuidzinas2012superconducting}, and will appear as an `unavoidable noise peak' in the DM mass region after Fourier transforming the readouts in the angular direction on this circular perimeter. Because the angle is very sensitive to frequency near resonance, this provides extreme precision in measuring the time stability of the gap.

If the noise from other sources is comparable to or smaller than the effect of the sinusoidal DM background, the DM effect is expected to show up as a peak at $f_\Phi \equiv M_\Phi/2\pi$ in the noise power spectrum. This means that as the noise profile gets continuously improved over time, one always has the possibility of encountering the universal oscillating background generated by DM. Depending on the frequency range of interest for a given experiment, this could be taken into account in the data analysis procedures.

One can define $\mathcal{R}$ as the ratio between the magnitude of DM modulation and other noises, and then obtain
\begin{equation}
\begin{split}
\mathcal{R} & \simeq    \frac{(\alpha_K/2)^2 \, \left(\eta_{\Delta,\text{max}}\right)^2 \, t_\text{obs}}{2S(f_\Phi)}
\\ & 
\simeq \left(\frac{\eta_{\Delta,\text{max}}}{10^{-14}}\right)^2 \left(\frac{\alpha_K}{0.5}\right)^2  \left(\frac{t_\text{obs}}{1 \text{ year}}\right)\left(\frac{10^{-22}\,\text{Hz}^{-1}}{S(f_\Phi)}\right) . \hspace{-5mm}
\label{eq:SNR}
\end{split}
\end{equation}
for a single resonator where $\alpha_K$ is the fraction of the kinetic inductance compared to the total inductance, $t_{\text{obs}}$ is the observation time, and $S(f_\Phi)$ is the noise power spectral density (PSD) at a given frequency $f_\Phi$. The detailed derivation of the formula is given in the Appendix.

Typical KID experiments are arranged in an array of $\mathcal{O}(10^3)$ resonators for a single readout chain \cite{golwala2024new,defrance2025low,hempel2026low,Golwala:2012dk,Smith:2024ljo}, and multiple readouts are operated depending on the experiment. If one has a set of total $N$ resonators in a coherent sinusoidal DM bath with uncorrelated noise backgrounds, the ratio would be multiplied by the factor of $N$; thus the sensitivity needed to achieve $\mathcal{R}\simeq1$ is also proportionally relaxed.

\begin{figure}[tb]
\centering
\includegraphics[width=0.49\textwidth]{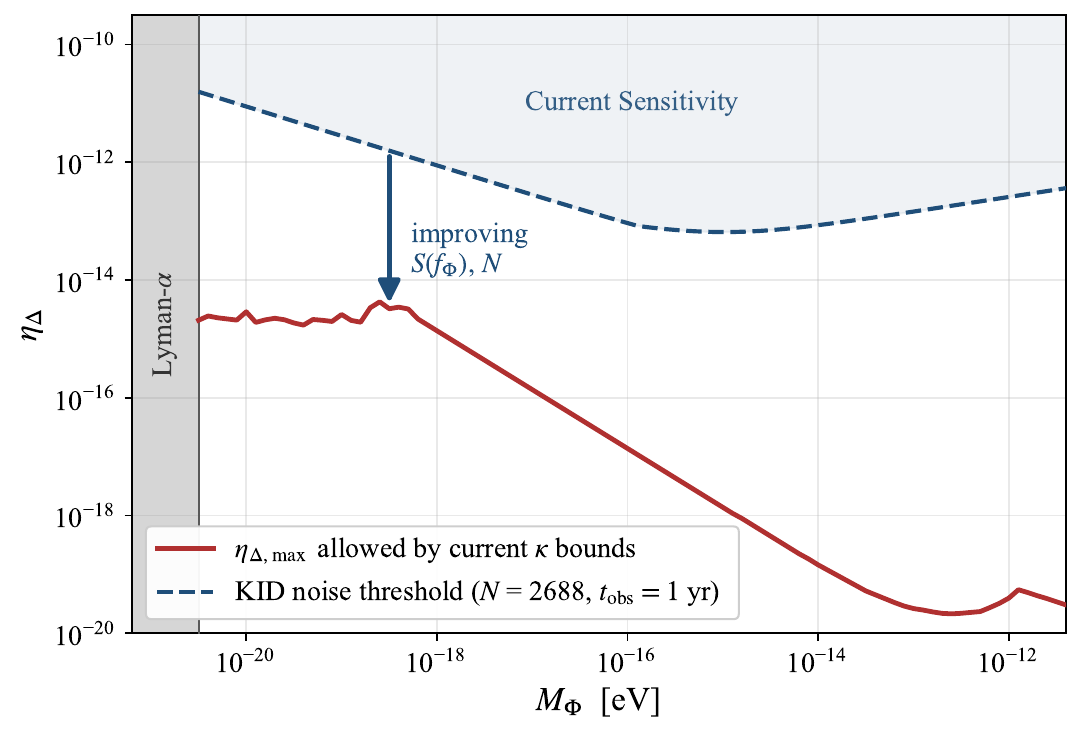}
\caption{ 
The maximum predicted fractional modulation size from the wave DM Yukawa coupling, depending on the DM mass. Current non-excluded regions from MICROSCOPE and atomic clock experiments yield the maximum allowed fractional modulations of $10^{-14}$ at DM mass below the mHz range. 
The dashed line represents the experimental sensitivity ($\mathcal{R}=1$) and is adapted from \cite{hempel2026low}, which proposes a total of 2688 Al resonators \cite{golwala2024new}.
}
\label{etaPlot}
\end{figure}

Figure~\ref{etaPlot} illustrates the comparison between recent experimental sensitivities and the maximum fractional modulations allowed under the current parameter constraints given by Eq. (\ref{eq:eta_max}), depending on the DM mass. 
Since the wave DM living in the surviving parameter space can result in $\sim \mathcal{O}(10^{-14})$ oscillations in the order parameter for ultralight DM scenarios, the improvement on the experimental side suggests that this intersection between the condensed matter side may be viable in future machinery.

\section{Summary and outlook}
\label{sec:summary}

In this work, we explored the intersections between particle and condensed matter physics, focusing on the implications of a wave DM background in superconducting systems. Because both areas are rapidly expanding, the direct interplay between ubiquitous wave DM and the electron sector may enter current theoretical schemes, leading to distinctive phenomena.

We demonstrate that the effective BCS Hamiltonian is modified by the Yukawa interaction to actually give the modulation of the macroscopic parameters. Our results show that through a feeble Yukawa coupling, the dark sector sources a time-periodic correction to the gap equation, transforming the superconducting order parameter into a dynamical degree of freedom that oscillates at the DM frequency.

This fundamental shift in perspective—from treating superconductors as passive engineered detectors to viewing them as cosmologically modified quantum systems—opens a novel connection between the $\Lambda$CDM picture and condensed matter physics. Building upon the theoretical discussion, we propose an explicit experimental setup illustrating how these microscopic background modulations may translate into actual frequency shifts in the resonator architecture.

Our results motivate several natural future directions. It is natural to examine how the DM with larger mass affects solid state materials, where the assumptions used in previous discussions might be altered. Our methodology assumes a small deviation from the adiabatic self-consistency condition in BCS superconductors, which can fail for sufficiently heavy DM. Once this slow-varying limit breaks down, its effect on condensed matter may change qualitatively, possibly making the theoretical prediction highly challenging and nontrivial. Nevertheless, this heavy mass region provides a rich landscape for future exploration, potentially revealing interesting physical phenomena such as Higgs-mode resonances \cite{shimano2020higgs}.

The material dependence can also be a crucial consideration for further analysis. Since the response coefficient $\mathcal{D}(x)$ depends on the choice of materials, different metals could give distinct gap response sizes to the same DM backgrounds. Also, while this study focuses on elements such as aluminum, which are simple and well suited for applying the small-coupling BCS framework and the free-electron model, other solid-state materials would violate some of the assumptions and might exhibit different modulation behaviors. Nevertheless, our theoretical framework can also be generalized to calculate the modulation effects in those materials, provided that one combines the suitable model with our framework. This universality and flexibility of our scheme make the exploration of various materials highly advantageous.

Furthermore, the dependence on the BSM model should also be a subsequent path to investigate. In this paper, we only considered the effect of Yukawa coupling with only electrons, but in principle the DM can also couple to hadrons and gauge bosons. This will modify the response of the phonon sector to the relic DM, eventually altering the scaling relation of $\omega_D$ and $\eta_\Delta$. Also, although our analysis focused on a simplified scalar wave DM model, this framework can be naturally extended to other theoretically motivated candidates such as axions or axion-like particles. In such scenarios, the Yukawa-type interaction would be modified to reflect the pseudoscalar nature of the axion. Given that we have demonstrated the feasibility of the effects of scalar wave DM in condensed matter, extending this approach to axions could open new avenues about the interplay between the beyond Standard Model phenomenology and the condensed matter physics.

\section*{Acknowledgements}
HL thanks B. Batell for helpful discussions, and P. Coloma, L. Molina-Bueno, and the other organizers of Light Dark World 2025 at IFT, Madrid, where part of this work was carried out. This work was supported in part by the National Research Foundation of Korea (Grant No. RS-2024-00352537). JY acknowledges support from the KAIST Jang Young Sil Fellow Program. JL is additionally supported by a KIAS Individual Grant (Grant No. QP110101) via the Quantum Universe Center at Korea Institute for Advanced Study.

\appendix

\section{Calculations of impacts on KID systems}

The kinetic inductance detector is arranged into an $LC$ circuit where superconducting materials act as circuit elements. Since an $LC$ circuit oscillates at its resonance frequency, probing it yields detailed information about its inductance value.

For a typical $LC$ circuit, the oscillation frequency $\nu$ is given by
\begin{equation}
\nu = \frac{1}{2\pi \sqrt{LC}},
\end{equation}
where $L$ is the inductance and $C$ is the capacitance in the circuit. For a superconducting material, the imaginary part of the conductivity is given as a function of the gap $\Delta$, and the total effective inductance $L$ of the object can be derived from its impedance.

The total inductance is written as the sum of a classical part depending on the geometry and a kinetic part generated by superconductivity~\cite{Mattis:1958yid}, $L = L_\text{geo} + L_K$.
In the limit of low temperature and slow resonant frequency compared to the energy gap, the kinetic inductance $L_K$ of the thin metal film can be expressed as a function of the gap $\Delta$~\cite{gao2008physics, zmuidzinas2012superconducting},  
\begin{equation}
L_K = 
\frac{\hbar R}{\pi \Delta} ,
\end{equation} 
where $R$ is the sheet resistance of the target material. This yields the fractional modulation of $L_K$ in terms of $\eta_\Delta$ as
\begin{equation}
\eta_{L_K} = -\eta_\Delta .
\end{equation}


$L_{\text{geo}}$ and $C$ may also depend on the electron mass in a nontrivial way in practical systems, but here we treat these to be fixed and focus on the shift propagated through the superconducting gap channel. Then, the above relation directly translates the gap modulation into the change in the resonance frequency as
\begin{equation}
\eta_\nu(t) \simeq \frac{1}{2}\alpha_K \eta_\Delta(t),
\end{equation}
where $\alpha_K \equiv L_K/(L_K+L_{\text{geo}})$ is the fraction of the kinetic inductance compared to the total inductance and $\eta_\nu(t)$ denotes the fractional deviation of the frequency $(\nu-\nu_0)/\nu_0$ from unperturbed resonance frequency $\nu_0$ of the KID. 

For sinusoidal perturbations submerged in other incoherent noise, the Fourier transform yields coherent time integration for this modulation, while other noise adds incoherently. This allows us to define a `signal-to-noise' like ratio of DM modulation and experimental noise spectra, increasing in total observation time, following the accumulation of data~\cite{Kahn:2016aff,Budker:2013hfa,Foster:2017hbq}
\begin{equation}
\mathcal{R} \equiv \frac{\text{(DM Modulation Power)}}{\text{(Exp. Noise Power)}} = \frac{\frac{1}{2}\eta_{\nu,\text{max}}^2  \, t}{S(f_\Phi)},
\end{equation}
where $t=t_{\rm obs}$ for the total observation time $t_{\rm obs}$ shorter than the coherent time of DM $t_{\rm coh} \sim 2\pi\times10^6/M_\Phi$, or $t={\sqrt{t_{\rm obs}t_{\rm coh}}}$ for a longer time. $\eta_{\nu,\text{max}}$ again represents the amplitude of the fractional modulation of the $LC$ resonance frequency.
The circuit frequency is modulated by the DM oscillation with $f_\Phi$.
$S(f_\Phi)$ denotes the power spectral density (PSD) function of the noise evaluated at the DM frequency $f_\Phi$.
As an estimate of experimental sensitivity, we regard the DM effect as potentially important when its spectral power exceeds the expected noise power in the relevant frequency bin, corresponding to $\mathcal{R}>1$, similar to the signal-to-noise comparisons commonly adopted in experimental analyses~\cite{Kahn:2016aff,Budker:2013hfa,Foster:2017hbq}.

The main noise sources in KID experiments include two-level system noise associated with dielectric, quasiparticle generation-recombination fluctuations, and standard amplifier noise.
Here, we perform numerical analysis on $\mathcal{R}$ considering the microstrip-coupled parallel-plate capacitor lumped-element KIDs~\cite{mazin2010thin, ji2014design, beldi2019high, cothard2024parallel, doyle2008lumped}, focusing on the proposed aluminum KID astronomical sensor~\cite{defrance2024characterization,martin2024hierarchical,golwala2024new,defrance2025low,hempel2026low}, 
which is a suitable choice for applying our model, since aluminum is a conventional low-temperature, weak-coupling type-I superconducting material.

Substituting the fractional gap modulation formula Eq. (\ref{eq:eta_max}) into the above one yields Eq. (\ref{eq:SNR}), with representative numbers coming from our selection of the benchmark points and the reference experiments.

For low modulation frequencies below kHz, which corresponds to ultralight DM with masses $M_\Phi \lesssim 10^{-12} \text{ eV}$, empirically, experimental demonstrations suggest that noise PSD can be assumed as a sum of $1/f$ flicker noise and white floor at these extremely low frequency regions~\cite{Burnett:2014lra,defrance2025low,hempel2026low}
\begin{equation}
S(f) \sim \frac{A}{f} + B,
\label{noisePSDform}
\end{equation} 
where $f$ is the fluctuation frequency.
$A$ and $B$ are numerical factors determined by geometry, experimental setup, and material properties. Given the long-time stability of these kinds of experiments, which are intended to operate continuously for years, we presume this formula for the KID array is applicable in this work for our range of interest in the DM mass.

Current state-of-the-art experiments \cite{hempel2026low} achieve $S\sim\mathcal{O}(10^{-18}\,\text{Hz}^{-1})$ for 1 Hz and above, while showing $S\sim\mathcal{O}(10^{-18}/f)$ at lower frequencies. This indicates that the noise PSD of $10^{-19}\,\text{Hz}^{-1}$ at the DM mass range could be viable in the future, entering the era where DM interactions with macroscopic solid state matter become crucial.


\bibliographystyle{elsarticle-num} 
\bibliography{ref}






\end{document}